# Harmonic generation by metal nanostructures optically coupled to 2D transition metal dichalcogenide

Elena Drobnyh[1], Ruth Pachter[2], Maxim Sukharev[1,3]

[1]Department of Physics, Arizona State University, Tempe, Arizona 85287, USA
[2]Materials and Manufacturing Directorate, Air Force Research Laboratory, Wright-Patterson Air Force Base, Dayton, Ohio 45433, USA
[3]College of Integrative Sciences and Arts, Arizona State University, Mesa, Arizona 85201, USA

Rigorous electrodynamical simulations based on the nonlinear Drude model are performed to investigate the influence of strong coupling on high harmonic generation by periodic metal gratings. It is shown that a thin dispersive material with a third order nonlinearity strongly coupled to surface plasmon-polaritons significantly affects even harmonics generated solely by the metal. The physical nature of this effect is explained using a simple analytical model and further supported by numerical simulations. Furthermore, the behavior of the second and third harmonics is investigated as a function of various physical parameters of the model material system, revealing highly complex dynamics. The nonlinear optical response of 2D few-layer WS$_2$ with both second and third order susceptibilities coupled to a periodic plasmonic grating is shown to have a significant effect on the second harmonic generation of the metal.

## INTRODUCTION

The field of plasmonics has enjoyed its rapid expansion into the molecular physics domain by combining unique optical properties of the surface plasmon resonance and highly spatially localized electromagnetic fields associated with strong coupling[1] to molecular aggregates, quantum dots,[2] and 2D materials[3] over a large scale of frequencies. Although much has been achieved in the linear optical regime,[4] a wide range of nonlinear optical phenomena such as second harmonic generation[5] and optical bistability[6] may bring new and exciting insight for application that are not investigated widely. There are just few works on exploring the large landscape of nonlinear plasmonics[7] combined with strong coupling to quantum systems,[8] while 2D materials, e.g. graphene, phosphorene, transition metal dichalcogenides (TMDs), group III and IV metal chalcogenides, as well as heterostructures such as graphene-BN, TMD-graphene and TMD-TMD, have been investigated for nonlinear optical applications,[9] including for broadband optical devices, nonlinear frequency conversion processes, mid-infrared photonics, and THz sources and detectors.

As TMDs in particular have shown much promise for photonics,[10-11] we chose tungsten disulfide (WS$_2$) as a model compound for few-layer 2D materials, for which these nonlinearities have been measured.[12-13] In our previous work[14] we studied WS$_2$ by a phenomenological third order susceptibility and discussed how plasmonic materials influence third harmonic generation, but the metal response in that work was considered only in the linear regime. Here we expanded both on the methodology, as well as on its application to a complex realistic model system. In this work the scope is twofold: first, we combine the nonlinear Drude model, adapted to periodic plasmonic systems, with a nonlinear response of a 2D material exhibiting instantaneous second and third order nonlinearities; and secondly, we explore the strong coupling regime and its influence on the second harmonic generation exhibited by the metal. The numerical results are supported by a simple analytical model of a driven anharmonic oscillator coupled to a Lorentz oscillator via dipole-dipole coupling.

## METHODS AND COMPUTATIONAL DETAILS

The electrodynamics of the plasmonic system under consideration is simulated by numerically integrating Maxwell's equations

$$\frac{\partial \vec{B}}{\partial t} = -\nabla \times \vec{E},$$
$$\varepsilon_0 \frac{\partial \vec{E}}{\partial t} = \frac{1}{\mu_0} \nabla \times \vec{B} - \vec{J}, \quad (1)$$

where $\varepsilon_0$ and $\mu_0$ are the permittivity and permeability of free space, respectively, and $\vec{J}$ is the current density in spatial regions occupied by metal. To account for the nonlinear optical response of the metal we adopt the classical theory based on the nonlinear Drude model.[15-17] The following set of equations is coupled to (1)

$$\partial_t n_e + \nabla(n_e \vec{u}) = 0,$$
$$m^*_e \left( \partial_t + (\vec{u} \cdot \nabla) + \gamma \right) \vec{u} = q_e (\vec{E} + \vec{u} \times \vec{B}), \quad (2)$$

where $n_e$ is the electron number density, $\vec{u}$ is the velocity field, $m^*_e = 0.99 m_e$ is the effective mass of conductive electrons for silver, and $\gamma$ is the phenomenological damping parameter. The current density in (1) is calculated via $\vec{J} = e n_e \vec{u}$.

It can be shown[15] by using (1) and (2) that the optical response of the metal in the linear regime is described by the Drude model

$$\varepsilon(\omega) = 1 - \frac{\Omega_p^2}{\omega^2 - i\omega\gamma}, \quad (3)$$

where $\Omega_p = \sqrt{e^2 n_0 / \varepsilon_0 m_e}$ is the plasma frequency. For silver, we used the following parameters: $\Omega_p = 9.04$ eV and $\gamma = 0.02125$ eV.

The Maxwell equations (1) are discretized in space and time along with the corresponding equations on the number density and the velocity field (2) in spatial regions occupied by the metal. The spatial discretization of eqs. (2) was found to be quite sensitive to our choice of whether velocity field components were collocated on the grid with the corresponding electric field components. We found that collocating the velocity field with the electric field using the Yee cell resulted in linear spectra identical to those we calculated using a conventional Drude model. On the other hand, if all the velocity field components were placed at the corners of the Yee cell the linear spectra were significantly different due to necessary spatial interpolations. Interestingly, the latter leads to highly distorted and noticeably lossy spectra. This however could be improved by employing the Lagrange polynomials interpolation,[18] which may come in handy in three-dimensional simulations. For our geometry though a simple collocation of the velocity field and the electric field was enough.

Additionally, the current density associated with the dispersion of WS$_2$ is included in the simulations in both the linear and nonlinear regimes. The resulting system of equations is propagated in time and space using the finite-difference time domain (FDTD) technique.[19] Numerical convergence is achieved for a spatial resolution of 1.0 nm and time step of $2 \times 10^{-3}$ fs. The open boundaries are terminated via convolutional perfectly matched layers.[20] The home-built codes are parallelized using the domain decomposition method and message passing interface (MPI).[21] All simulations are performed at our local multiprocessor cluster. A typical number of computing cores used for a single run is 64, with an average execution time of 10-20 minutes. To probe the system under consideration we employ the total field/scattered field approach.[22]

Linear spectra are simulated using the short pulse method, which allows obtaining reflection/transmission spectra in a single FDTD run.[23] The nonlinear power spectra are simulated using a 100 fs incident pulse. We found that it is notoriously difficult to achieve a steady-state regime under strong coupling conditions (i.e. when WS$_2$ is placed very close to the metal) when pumped at one of the polariton frequencies. In many cases the time propagation for more than 1 ps resulted in numerical divergence. Therefore, the pulse excitation scheme was chosen. We tested our main results with respect to the incident pulse duration and found that they were not dependent on the pulse duration so long as it was 50 fs or longer. It is interesting to note that in our previous studies when the nonlinear response of the metal was not considered[14] we did not encounter any numerical issues when obtaining steady-state solutions. We attribute the observed numerical divergence to our choice of collocation of the velocity field and the electric field on the grid.

**RESULTS AND DISCUSSION**

First we discuss the model system considered and its linear optical response. The plasmonic system we considered consists of a flat metal film of thickness *d* placed on the glass substrate and metal wires of width $R_x$ and height $R_y$ on top of the flat metal film (see Fig. 1a). The experimental implementation of such a geometry usually contains a few nm thin dielectric spacer between the film and wires,[24] allowing one to combine different metals. We verified that the spacer has no effect on the electromagnetic properties of the system and it is not considered in our simulations. Moreover, to avoid lightning rod effects due to sharp corners, we used smooth corners of the wire with curvature *R*, as also schematically depicted in Fig. 1a. To investigate the influence of the thin nonlinear optical material coupled to the plasmonic response, it is placed on the input side at a distance *L* above the metal wire. Although the glass substrate is considered in our simulations, we found that it has a very little effect when the thickness of the metal film exceeds 100 nm.

When investigating the dynamics of plasmons coupled to a thin dispersive material (schematically depicted as WS$_2$ in Fig. 1a), the latter is described by the linear Lorentz dielectric function with experimental parameters corresponding to monolayer WS$_2$.[25] The nonlinear part of the response is simulated by adding second-, $\chi^{(2)}$, and third-order susceptibilities, $\chi^{(3)}$, and implementing an efficient numerical procedure allowing to account for both linear dispersion and the nonlinear effects.[26] We note that the explicit numerical technique[26] allows us to easily turn on and off any nonlinear susceptibility, which helps to investigate various nonlinear regimes and their influence on high harmonic generation by the metal. The numerical parameters used in our simulations correspond to nonlinear experimental measurements for WS$_2$: $\chi^{(2)} = 10^{-9}$ m/V and $\chi^{(3)} = 2.611 \times 10^{-17}$ m$^2$/V$^2$.[12-13]

The main linear optical properties of the periodic plasmonic grating under consideration are summarized in Figs.1b-d. We note that the transmission for a film thickness of 100 nm is nearly 0, making the reflection a reliable observable to identify plasmon resonances of the system. For the geometrical parameters considered the reflection has two well pronounced minima. A sharp high energy mode (not shown in Fig. 1) corresponds to the electromagnetic anomaly associated with the periodicity of the grating. The surface plasmon-polariton modes (low energy minima in Figs. 1b and 1c) are highly dependent on the width of the metal wire, $R_x$, (panel (b)) and also on the periodicity (panel (c)).

For a fixed value of wire width $R_x$, a decrease in the period leads to a predictable blue shift of the mode as the wires are closer, as shown in Fig. 1c. We note that the efficiency of the coupling to the plasmon mode (defined as the absorption, which in our case is simply 1-reflection) also varies with both the period and the wire's width. Fig. 1d shows the dependence of the reflection and the corresponding plasmon frequency on $R_x$ for a period value of 550 nm. When the value of $R_x$ is close to that of the period the reflection reaches nearly 35%, demonstrating that the coupling of the incident light to the plasmon mode is very efficient. Increasing $R_x$ results in a red shift of the plasmon resonance with the reflection (Fig. 1b), going from merely 90% to 35% efficiency. One may use $R_x$ as a tuning knob to adjust the position of the mode to a desired frequency, which is very useful when coupling to another optical oscillator such as a molecular aggregate[1] or a 2D material.[14]

For large values of $R_x$ (i.e. small gaps between wires), the plasmon mode becomes highly spatially localized near/inside the gaps between wires. Although the reflection becomes significantly smaller in this case, such localization is not desirable since our goal is to be able to couple the plasmon mode to a 2D material placed in proximity to the wires. Thus, we keep $R_x$ large enough for the incident field to couple to the plasmon, but small enough such that the plasmon mode spatially extends to about 15-50 nm above the wire's surface.

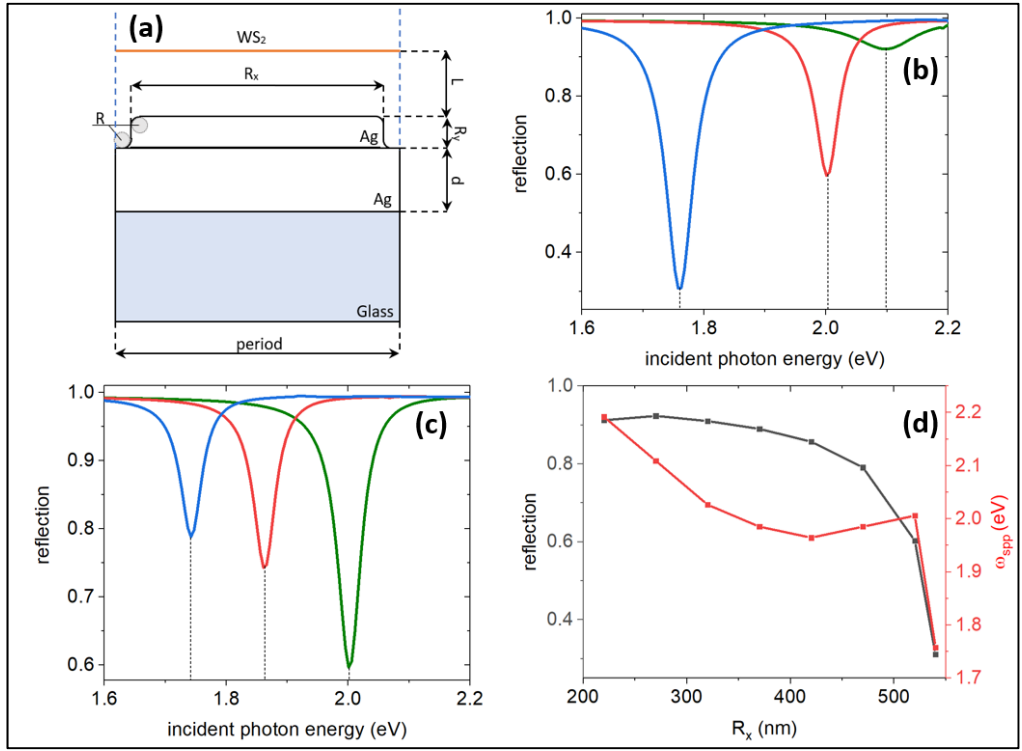

**Fig. 1.** Linear optics of periodic arrays of wires on a metal film. Panel (a) shows the schematic setup of the system comprised of the metal wire of height $R_y$ and width $R_x$, flat metal film, the glass substrate, and 2D dispersive WS$_2$ material. A few-layer thin WS$_2$ is placed at a distance $L$ above the metal. In all simulations the thickness of the silver film, $d$, is 100 nm, the wire's height, $R_y$, is 50 nm, and the curvature of the corners, $R$, is 10 nm. Both panels (b) and (c) show the reflection spectra of the system without WS$_2$ as a function of the incident photon energy. In panel (b) different colors correspond to different values of the wire's width $R_x$ keeping other parameters constant. The period is set at 550 nm. The blue line corresponds to $R_x = 540$ nm, the red line is for $R_x = 520$ nm, and the green line is for $R_x = 270$ nm. In panel (c) different colors correspond to different values of the period and $R_x = 520$ nm. The blue line is for a period of 650 nm, the red line corresponds to a period of 600 nm, and the green line to a period of 550 nm. Panel (d) shows the reflection coefficient evaluated at the corresponding plasmon frequency as a function of $R_x$ (left vertical axis, black line) and the plasmon frequency (in eV), as a function of $R_x$ (right vertical axis, red line).

The nonlinear response of the bare plasmonic system without the 2D material is discussed next (see Fig. 2), specifically the dependencies of the second (SH) and third (TH) harmonics on various geometrical parameters of the model system. The frequency resolved signal obtained from 100 fs intense pulse excitation is shown in Fig. 2a at three incident pump amplitudes. The SH is clearly visible for all pump intensities, while the TH signal appears to show at $10^8$ V/m. A tincture of the fourth harmonic is also seen for this pump amplitude. We note that all harmonics are noticeably broad with non-symmetric shapes, indicating that interference effects between waves generated by different parts of the system may play an important role. To check the numerical convergence of our model and to ensure that peaks exhibited in Fig. 2a correspond to harmonics generation we integrate SH and TH signals over frequencies to account for losses and dispersion and plot them as functions of the pump intensity in Fig. 2b. The slopes of both SH and TH are nearly the ideal 2 and 3 values expected, respectively.

Next, we explore the influence of structural parameters and the effect of the plasmon resonance on the SH (Fig. 2c) and TH (Fig. 2d) signals. Two scenarios are considered: (a) we fix the pump frequency at 2 eV (corresponds to the plasmon resonance for $R_x$ of 520 nm) and vary $R_x$ (black lines in Figs. 2c and 2d); (b) when varying $R_x$ we adjust the pump frequency to the given geometry (red lines). It is expected that when pumped at the plasmon resonance the efficiency of SH and TH generation would be prominently higher due to local electromagnetic field enhancement. However, our results suggest more complex behavior of the SH generation process. As seen from Fig. 2c the SH signal exhibits two maxima for a fixed pump frequency: (1) an obvious enhancement at $R_x$ = 520 nm as this corresponds to the resonant conditions; (2) an SH signal for $R_x$ = 320 nm is observed almost as high as in case (1). When comparing SH signals off (black) and on resonance (red) both conditions result in nearly the same SH efficiency for values of $R_x$ above 300 nm. However, the SH signal drops almost by an order of magnitude for narrow wires (small $R_x$). The TH signal (Fig. 2d) depends on $R_x$ as one expects with the signal nearly monotonically increasing with $R_x$.

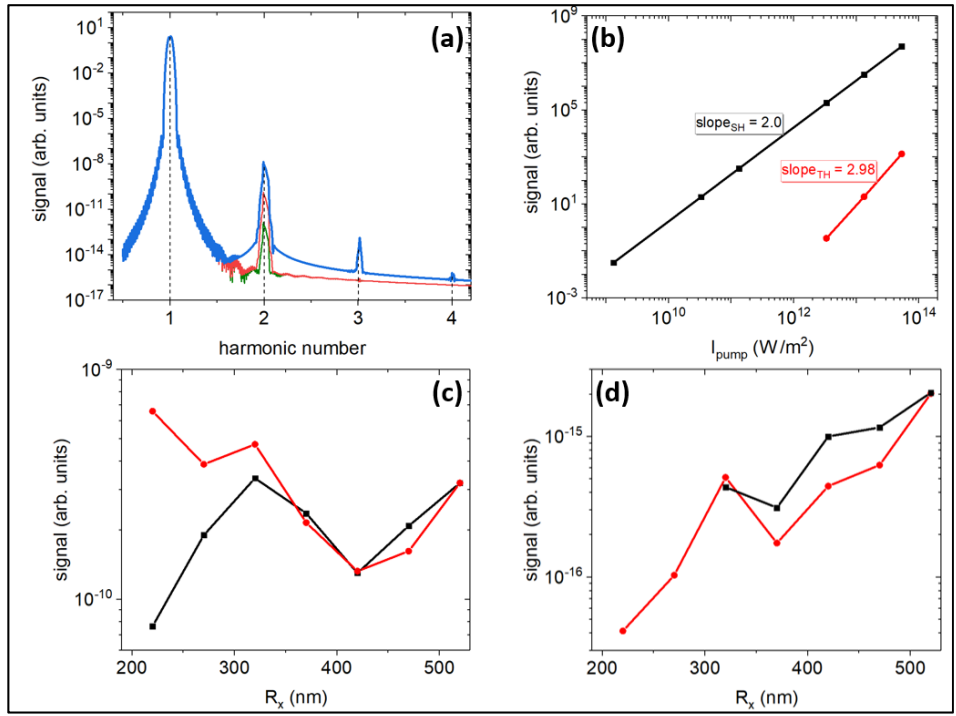

**Fig. 2**. Nonlinear optics of periodic arrays of wires on a metal film. Panel (a) shows the harmonics spectra evaluated at three different pump amplitudes: $E_0 = 10^6$ V/m (green line), $E_0 = 10^7$ V/m (red line), and $E_0 = 10^8$ V/m (blue line). The pump frequency is 2 eV corresponding to the plasmon mode for the period of 550 nm and wire's width $R_x = 520$ nm. The signal is normalized with respect to the pump intensity. Panel (b) shows the signal evaluated for the second (black line) and the third (red line) harmonics as function of the pump intensity for the same geometrical parameters as in panel (a). Both panels (c) and (d) show the second and the third harmonic signals, respectively, as function of $R_x$ when pumped at the fixed plasmon resonance corresponding to $R_x = 520$ nm (black line) and when pumped at the resonant frequency evaluated for each value of $R_x$ (red line).

Following analysis of high harmonic generation by the plasmonic system, we now turn to a hybrid construct combining the plasmonic grating with a thin 2D material placed on the input side (see Fig. 1a). In order to quantitatively explore the nonlinear optical response of such a system we use the Lorentz model for the 2D material with parameters that describe WS$_2$. Fig. 3a shows the linear reflection spectra of the hybrid system with WS$_2$ placed at 15 nm above the metal wires (L in Fig. 1a). By adjusting geometrical parameters of the grating, we find the optimal geometry when the frequency of the plasmon mode matches that of the WS$_2$ direct band-gap exciton. The goal of simulations shown in Fig. 3 is to understand how a dispersive material may affect even harmonics generation (solely produced by the metal) in the strong coupling regime. The strong coupling regime between the surface plasmon mode supported by the grating and WS$_2$ is indeed achieved. In order to confirm that the splitting noted in Figure 3a for the period of 550 nm and $R_x = 520$ nm is indeed due to exciton-plasmon coupling, we perform a set of simulations, varying the period of the grating and tracing frequencies of the upper and lower polaritons. We observe a clear avoided crossing between the exciton and the plasmon mode, ensuring that the system is in the strong coupling regime.

Interestingly, the harmonics spectra of a bare plasmonic grating and the hybrid system (Fig. 3b), clearly demonstrate two important effects: (1) odd harmonics are significantly enhanced by WS$_2$; (2) even harmonics are greatly affected by the presence of WS$_2$ even though the model does not directly include the second order susceptibility for WS$_2$. The latter indicates that WS$_2$ influences even harmonics that are otherwise generated exclusively by the metal.

In order to further elucidate and understand this effect, we consider a simple model of two coupled oscillators with one being externally driven and having a second order nonlinearity

$$\ddot{x}_1 + \gamma_1 \dot{x}_1 + \omega_1^2 x_1 + a x_1^2 - g^2 x_2 = F_0,$$
$$\ddot{x}_2 + \gamma_2 \dot{x}_2 + \omega_2^2 x_2 - g^2 x_1 = 0, \tag{4}$$

where $x_{1,2}$ correspond to the displacements of the two oscillators coupled through the dipole-dipole coupling proportional to the parameter $g^2$. The first oscillator is driven by the external force $F_0$ and has a second order nonlinearity characterized by the parameter $a$. To show the analogy with the hybrid system we note that the first oscillator represents the metal and the second one is the 2D material. Since the extinction of WS$_2$ is significantly smaller compared to that of the metal we assume that only the metal is driven by the incident field.[27] Such a consideration is valid in the near-field when the exciton-plasmon coupling is the major contributor to the dynamics. We thus neglect the filtering effect, which dominates at large distances when the coupling is small. When the incident field is small, the second order nonlinearity in the first equation is significantly smaller compared to other terms, and we can apply the Rayleigh-Schrödinger perturbation theory[28] to find the second order contribution $x_1^{(2)}$ for the process of the second harmonic generation

$$x_1^{(2)} = -\frac{aF_0^2}{\omega_1^2 + i\gamma_1(2\omega) - (2\omega)^2} \frac{\left(\omega_2^2 + i\gamma_2\omega - \omega^2\right)^2}{\left(g^4 - \left(\omega_1^2 + i\gamma_1\omega - \omega^2\right)\left(\omega_2^2 + i\gamma_2\omega - \omega^2\right)\right)^2}. \tag{5}$$

When the coupling between two oscillators, $g$, is negligible, $x_1^{(2)}$ reduces to the well-known expression[28]

$$x_1^{(2)} \approx -\frac{aF_0^2}{\left(\omega_1^2 + i\gamma_1(2\omega) - (2\omega)^2\right)\left(\omega_1^2 + i\gamma_1\omega - \omega^2\right)^2}. \tag{6}$$

Eq. (5) evidently shows that the second harmonic is significantly influenced by the presence of another dispersive material (even though this material does not have its own second order nonlinearity) as long as the coupling is strong enough. The corresponding second order susceptibility associated with (5) peaks at two well-defined frequencies

$$\omega_\pm^2 = \frac{\omega_1^2 + \omega_2^2}{2} \pm \sqrt{\frac{\left(\omega_1^2 - \omega_2^2\right)^2}{4} + g^4} \tag{7}$$

corresponding to the upper and lower polaritonic states of two coupled oscillators.[29]

We now return to our numerical model and explore how the SH signal varies with the distance between $WS_2$ and the plasmonic array. Fig. 3c shows our main results. Here we consider pumping the system using the frequency corresponding to either the lower or the upper polaritonic states (see the resonance splitting in Fig. 3a, blue line). The SH signal for the lower polariton pump drops as a function of the distance $L$ as expected, while the pumping at the upper polariton leads to a surprising increase of the SH signal with the distance between $WS_2$ and metal. We first note that the SH signal is higher when $WS_2$ is absent (see Fig. 3b, red vs black lines for the second harmonic) and one may expect that both lines in Fig. 3c need to reach an asymptotic value at very large distances $L$. However, in our simulations this does not happen because even though at large values of $L$ $WS_2$ is not directly coupled to the surface plasmons it still influences the electromagnetic field that is driving the plasmonic grating. Since $WS_2$ is located on the input side, a filtering effect results. We performed test simulations placing $WS_2$ as far as 1 µm away from the array and found that the SH signals begin to level out as a function of $L$ after about 100 nm, confirming the filtering effect at large distances. The analytical model presented above can be used to explain the striking difference between different pumping frequencies for the SH signal seen in Fig. 3c.

Fig. 3d shows the real part of the second order contribution $x_1^{(2)}$ for the case when the eigenfrequencies of each oscillator are slightly displaced relative to one another. When the coupling strength increases (i.e. $WS_2$ is placed closer to the plasmonic array) the hybrid states are clearly visible. The SH signal is significantly non-symmetric, exhibiting higher values at the lower polaritonic branch since two oscillators have both different damping parameters (as in the case of our real system, where the metal is significantly more lossy than the exciton mode of $WS_2$) and different eigenfrequencies. Thus, the SH signal behaves differently as a function of the coupling strength, $g$, depending on whether the pump frequency cuts through the lower or the upper polaritons.

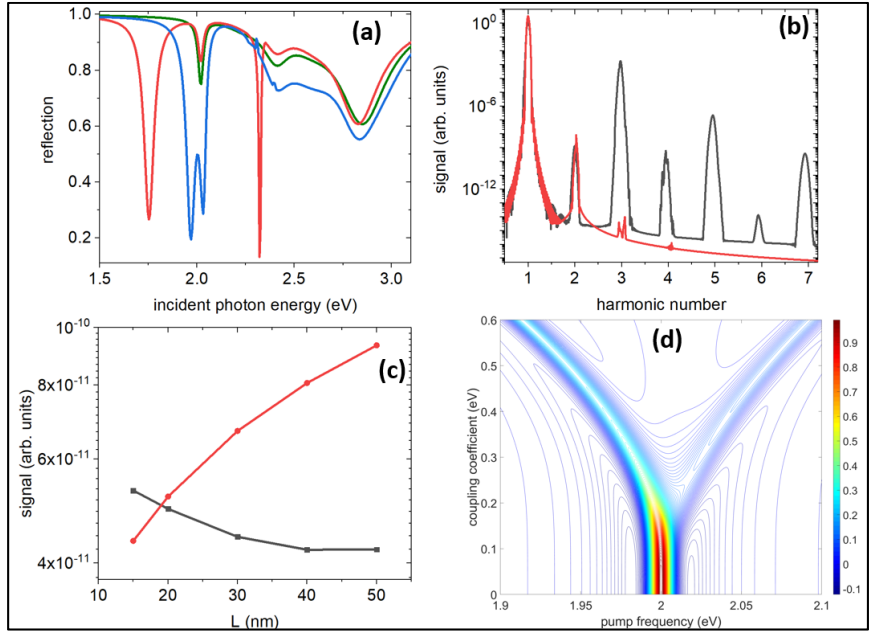

**Fig. 3**. Nonlinear optics of periodic arrays of wires on metal film strongly coupled to a 2D material with a third order nonlinearity. Panel (a) shows linear reflection spectra for the standalone 2D WS$_2$ (green line), for the WS$_2$ placed 15 nm above the grating for a period of 530 nm (red line) and a period of 550 nm (blue line). Panel (b) shows the harmonics spectra with (black line) and without (red line) WS$_2$ at the distance of 15 nm above the grating. The pump amplitude $E_0 = 10^8$ V/m, the pump frequency is 1.97 eV, period of the grating is 550 nm, and $R_x$ = 520 nm. Panel (c) shows the second harmonic signal as a function of the distance between the grating and WS$_2$, $L$, (see Fig. 1a) when pumped at the lower (black lines) and upper (red lines) polariton energies (1.97 eV and 2.03 eV), respectively. Panel (d) shows the normalized real part of the second order contribution, $x_1^{(2)}$, obtained from eq. (5) as a function of the pump frequency and the coupling parameter $g$. The following parameters for the oscillators were used: $\omega_1 = 2.0$ eV and $\gamma_1 = 2.13 \times 10^{-2}$ eV (plasmon), $\omega_2 = 2.01$ eV and $\gamma_2 = 2.8 \times 10^{-2}$ eV (exciton).

Monolayer WS$_2$ is non-centrosymmetric and exhibits a second order susceptibility, $\chi^{(2)}$, which was shown experimentally to reach remarkably high values due to large density of states.[12] Fig. 4 shows the second harmonic signal as a function of the pump frequency comparing three models: the periodic plasmonic grating only, the plasmonic grating with WS$_2$ taking into account only its $\chi^{(3)}$, and the plasmonic grating coupled to WS$_2$ with both $\chi^{(2)}$ and $\chi^{(3)}$. We note that the bare metal, as expected, generates the second harmonic more efficiently if pumped at the corresponding plasmon frequency, as it was pointed out earlier. When the second order nonlinearity of WS$_2$ is included in the simulations, the maximum of the second harmonic signal is redshifted to 1.94 eV, which is not the exciton absorption peak of bare WS$_2$ (2.01eV) but rather the lower polariton state of the coupled system. Obviously WS$_2$ dominates the process of the second harmonic generation due to a high value of $\chi^{(2)}$, but the spectrum does have a signature of the plasmonic material – the observed frequency shift and the linewidth of the peak is clearly broader than for a bare WS$_2$. When $\chi^{(2)}$ is neglected (or is small compared to the second order processes generated by metal) the second harmonic signal evidently has a signature of the strong coupling with two well-seen resonant peaks corresponding to the upper and lower polaritonic modes. It should be noted that the frequency shift observed in our calculations has been experimentally observed for a system comprised of c-porphyrin and a metal cavity.[8]

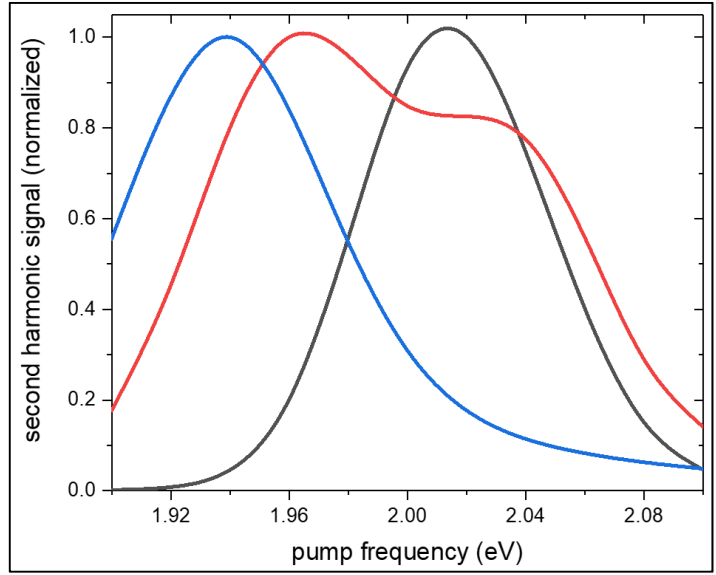

**Fig. 4**. Second harmonic signal (normalized) as a function of the pump frequency calculated for the plasmonic grating only (black line), the grating coupled to 2D material with only the third order nonlinearity (red line), and the grating coupled to 2D material with both second and third order susceptibilities (blue line). The second order susceptibility for the 2D material is $\chi^{(2)} = 10^{-9}$ m/V. Other parameters: the period of the plasmonic grating is 550 nm, $R_x = 520$ nm, $WS_2$-to-metal distance is 15 nm, the pump amplitude is $10^8$ V/m.

## CONCLUSION

In summary, we presented self-consistent calculations combining the nonlinear optical response of a periodic plasmonic system coupled to a thin nonlinear material exhibiting instantaneous second and third order nonlinearities (with parameters corresponding to tungsten disulfide). We demonstrated that under strong coupling conditions the second harmonic signal solely generated by the metal is significantly influenced by the dispersion of $WS_2$ even when the second order nonlinearity for $WS_2$ was neglected. A simple analytical model of two coupled oscillators was presented to explain this effect. We note that the results of our work can also explain recent experimental observations,[8] where significant enhancement and spectral distortion of the second harmonic peak was observed for strongly coupled c-porphyrin and a silver nanocavity. An obvious extension of this work is to explore high harmonic generation in exciton-plasmon systems in three dimensions, including not-yet discussed chiral properties of the second harmonic and their influence on the strong coupling.

## ACKNOWLEDGEMENTS

This work is sponsored by the Air Force Office of Scientific Research under Grant No. FA9550-19-1-0009. MS is also grateful to the generous financial support from the Binational Science Foundation through Grant N0. 2014113. The authors would like to thank Prof. Abraham Nitzan and Prof. Adi Salomon for stimulated discussion of the main results of this work.## REFERENCES